\documentclass[12pt]{article}
\usepackage{epsfig,latexsym,amsmath,amsfonts,mathabx,xcolor}

\usepackage{graphicx}
\usepackage{array}
\newcolumntype{L}[1]{>{\raggedright\let\newline\\\arraybackslash\hspace{0pt}}m{#1}}
\newcolumntype{C}[1]{>{\centering\let\newline\\\arraybackslash\hspace{0pt}}m{#1}}
\newcolumntype{R}[1]{>{\raggedleft\let\newline\\\arraybackslash\hspace{0pt}}m{#1}}

\newcommand{\bass}{\begin{assumption} \rm}
\newcommand{\eass}{\end{assumption}}
\newcommand{\bconj}{\begin{conjecture} \rm}
\newcommand{\econj}{\end{conjecture}}

\newcommand{\bthm}[1]{\begin{theorem}\index{theorem, #1}\index{#1} }
\newcommand{\ethm}{\qed \end{theorem} }
\newcommand{\blem}[1]{\begin{lemma}\index{lemma, #1}\index{#1} }
\newcommand{\elem}{\qed \end{lemma} }
\newcommand{\bcor}[1]{\begin{corollary}\index{corollary, #1}\index{#1} }
\newcommand{\ecor}{\qed \end{corollary} }
\begin{document}
\bibliographystyle{plain}
\newcommand{\widespace}{\hspace*{1em}}
\newcommand{\negspace}{\hspace*{-0.5em}}
\newcommand{\ack}{\paragraph{Acknowledgment: } \rm}
\newcommand{\instance}{\paragraph{Instance: } }
\newcommand{\reals}{{\mathbb R}}
\newcommand{\integers}{{\mathbb Z}}
\newcommand{\rationals}{{\mathbb Q}}
\newcommand{\discrete}{{\mathbb D}}
\newcommand{\complex}{{\mathbb C}}
\newcommand{\buses}{{\mathbb K}}
\newcommand{\lines}{{\mathbb E}}
\newcommand{\tree}{{\mathbb T}}
\newcommand{\graph}{{\mathbb G}}
\newcommand{\N}[1]{\mbox{${\cal N}_{\!{#1}}$}}  
\newcommand{\feasset}{{\mathbb S}}
\newcommand{\ternaryfeasset}{{\mathbb T}}
\newcommand{\noSOCfeasset}{{\mathbb M}}
\newcommand{\pumpfeasset}{{\mathbb P}}
\newcommand{\genfeasset}{{\mathbb G}}
\newcommand{\otherset}{{\mathbb P}}
\newcommand{\lattice}{{\mathbb L}}
\newcommand{\convex}{{\cal C}}
\newcommand{\map}{{\cal M}}
\newcommand{\vertices}{{\cal V}}

\newcommand{\pumpvar}{{p}}
\newcommand{\genvar}{{g}}
\newcommand{\Pumpvar}{{P}}
\newcommand{\Genvar}{{G}}
\newcommand{\statevar}{{s}}
\newcommand{\Statevar}{{S}}
\newcommand{\pumpcvar}{{u}}
\newcommand{\gencvar}{{v}}
\newcommand{\discretevar}{{z}}
\newcommand{\discretepumpvar}{{q}}
\newcommand{\discretegenvar}{{h}}
\newcommand{\shiftedstatevar}{{w}}
\newcommand{\epoch}{{e}}
\newcommand{\uneven}{{t}}
\newcommand{\identity}{{\bf I}}
\newcommand{\zero}{{\bf 0}}
\newcommand{\one}{{\bf 1}}
\newcommand{\sqrtmone}{\mbox{j}}

%
%
\newcommand{\trace}[1]{\mbox{\rm tr$(#1)$}} 
\newcommand{\diag}[1]{\mbox{\rm diag$\left(#1\right)$}} 
\newcommand{\topp}[1]{\left\lceil {#1} \right\rceil} 
\newcommand{\bott}[1]{\left\lfloor {#1} \right\rfloor} 
\newcommand{\stopp}[1]{\lceil {#1} \rceil} 
\newcommand{\sbott}[1]{\lfloor {#1} \rfloor} 
\newcommand{\round}[1]{\left[ {#1} \right]} 
%
%
\newcommand{\eval}[2]{\left. {#1}
        \raisebox{0ex}[0ex][4ex]{}
                \right|_{#2}}
\newcommand{\squared}[1]{({#1})^{\two}}
\newcommand{\cubed}[1]{( {#1} )^{\three}}
\newcommand{\abssquared}[1]{\left| {#1} \right|^{\two}}
\newcommand{\abs}[1]{\left| {#1} \right|}
\newcommand{\normsquared}[1]{\left\| \raisebox{-5ex}[8ex][0ex]{#1}
                    \right\|^{\two}}
\newcommand{\norm}[1]{\left\| {#1} \right\|}
\newcommand{\inverse}[1]{{#1}^{\!-\!1}}
\newcommand{\transpose}[1]{{#1}^{\dagger}}
\newcommand{\hermitian}[1]{{#1}^{\ddagger}}
\newcommand{\mtranspose}[1]{{#1}^{\!-\!T}\!}
%
\newcommand{\eb}{\begin{equation}}
\newcommand{\eab}{\begin{eqnarray}}
\newcommand{\ee}{\end{equation}}
\newcommand{\eae}{\end{eqnarray}}
\newcommand{\non}{\nonumber}
\newcommand{\eabn}{\begin{eqnarray*}}
\newcommand{\eaen}{\end{eqnarray*}}
%
%
%

%
\newcommand{\equal}[1]{\stackrel{[{#1}]}{=}}
\newcommand{\approxi}[2]{{#1}_{[{#2}]}}
%
\newcommand{\squish}{\renewcommand{\arraystretch}{1.05}}
\newcommand{\supersquish}{\renewcommand{\arraystretch}{0.75}}
\newcommand{\unsquish}{\renewcommand{\arraystretch}{1}}
\newcommand{\barr}{\squish \begin{array} }
\newcommand{\btarr}{\supersquish \begin{array} }
\newcommand{\bsarr}{\supersquish \scriptsize \begin{array}{@{}c@{}} }
\newcommand{\earr}{\end{array} \unsquish}
%
%
\newcommand{\dsum}{\displaystyle \sum}
\newcommand{\cross}{{\displaystyle \raisebox{-0.6ex}{\Large\sf X}}}
%
\newcommand{\ffrac}[2]{\barr{@{}c@{}}
{#1} \\ \hline {#2} \earr \,}
\newcommand{\sfrac}[2]{{\textstyle\frac{#1}{#2}}}
\newcommand{\partialdbyd}[2]{\barr{@{}l@{}}
\,\partial {#1} \\ \hline \partial{#2} \earr \,}
\newcommand{\totaldbyd}[2]{\barr{@{}l@{}}
\,d {#1} \\ \hline d {#2} \earr \,}
\newcommand{\partialdsbyds}[2]{\barr{@{}l@{}}
\,\partial^{2} {#1} \\ \hline \partial{#2}^{2} \earr\,}
\newcommand{\totaldsbyds}[2]{\barr{@{}l@{}}
\,d^{2} {#1} \\ \hline d {#2}^{2} \earr \,}
\newcommand{\partialdsbydsds}[3]{\barr{@{}l@{}}
\,\partial^{2} {#1} \\
\hline \partial{#2} \partial{#3} \earr}
\newcommand{\partialdcbydsdsds}[4]{\barr{@{}l@{}}
\,\partial^{3} {#1} \\
\hline \partial{#2} \partial{#3} \partial{#4} \earr}
\newcommand{\partialdcbydsds}[3]{\barr{@{}l@{}}
\,\partial^{3} {#1} \\
\hline \partial{#2}^{2} \partial{#3} \earr}
\newcommand{\partialdcbydc}[2]{\barr{@{}l@{}}
\,\partial^{3} {#1} \\
\hline \partial{#2}^{3} \earr}
\newcommand{\totaldsbydsds}[3]{\barr{@{}l@{}}
\,d^{2} {#1} \\
\hline d{#2} d{#3} \earr}
%
\newcommand{\bmat}[1]{\left[ \barr{{#1}} }
\newcommand{\emat}{\earr \right]}
\newcommand{\bvec}{\left[ \barr{r} }
\newcommand{\cvec}{\left[ \barr{c} }
\newcommand{\lvec}{\left[ \barr{l} }
\newcommand{\evec}{\earr \right]}
%
%
\newcommand{\players}{{\mathbb K}}
\newcommand{\strat}{{\mathbb S}}
\newcommand{\comp}{{\mathbb C}}
\newcommand{\activeset}{{\mathbb A}}
\newcommand{\depend}{{\mathbb D}}
\newcommand{\inmat}{A}
\newcommand{\eqmat}{F}
\newcommand{\new}{ }
\newcommand{\change}{\Delta\!}

\title{A state-of-charge based formulation for storage participation in electricity markets: \\ Technical Reference}

\author{Ross~Baldick \\ \small Department of Electrical and Computer Engineering \\ \small The University of Texas at Austin}

\markboth{\mbox{  }}%
{Shell \MakeLowercase{\textit{et al.}}: Bare Demo of IEEEtran.cls for Journals}

\maketitle
\begin{abstract}
We consider a storage device and develop a basic market design that represents the salient characteristics of storage such as state-of-charge (SOC) and round-trip efficiency. The contribution of this paper is a market design that reflects these technical characteristics, does not require bids and offers by the storage owner within the market horizon, but does require an end-of-horizon bid/offer for deviating the end-of-horizon SOC from the start-of-horizon SOC\@. Small examples are used to illustrate the market design and large-scale implementation is considered. Several extensions are sketched in Appendices.
\end{abstract}

\section{Introduction} \label{introsect}
With increasing levels of renewable resources in many electricity systems, and simultaneously a decreasing proportion of dispatchable thermal resources, significant amounts of storage capacity have been added recently in systems worldwide~\cite{JPM25,EIA25}. The trend of increasing installed capacity of energy storage systems (ESS) is anticipated to continue and even accelerate, particularly as the cost of storage continues to drop and as dispatchable thermal resources are retired. This paper responds to that trend by developing a proposal for tailoring the design of electricity markets to accommodate high levels of storage associated with high renewable penetration and decreasing availability of dispatchable thermal generation.

The approach in this paper departs from one current approach in, for example, United States (US) organized markets operated by Independent System Operators (ISOs) that represents storage as a paired (and possibly unified) generator and load, with offers and bids, respectively, applying to individual intervals over the dispatch horizon. Instead, the formulation here focuses on central dispatch of ESS based on the ISO optmimizing SOC while explicitly considering the technical characteristics of storage including round-trip efficiency, and with only a single bid/offer for deviating the end-of-horizon SOC from the start-of-horizon SOC\@.\footnote{We will slightly modify this statement in the light of self-discharge.} The formulation here follows, for example, \cite{Bal21,CAISO20}\cite{SiE20}\cite[Section II]{ZZL20}. 

This paper will focus on the market design relating to power and energy, whereas a future paper will consider ancillary services (AS) in detail. The contribution of this paper is a market design that reflects the technical characteristics of storage devices and does not require forecasts of prices by the ESS owner within the market horizon but does require an end-of-horizon bid/offer for deviating the end-of-horizon SOC from the start-of-horizon SOC\@. 

The organization of this paper is as follows.  Section~\ref{hispersp} provides a historical perspective on organized markets in the US and presents a philosophy for the way in which US electricity markets were designed for resource mixes that included mostly thermal resources. While there are several ways that such market designs could be expanded to represent ESS, Section~\ref{adapelect} will summarize the current approach in several US Regional Transmission Organization/Independent System Operator (RTO/ISO) markets and provides a critique in the context of the philosophy from Section~\ref{hispersp}. In particular, it will be argued that the current approach in US electricity markets of relying on ESS owners forecasting prices within the dispatch horizon to inform the offers and bids will be increasingly unmanageable because it will inherently result in ESS dispatch sub-optimality, and resulting higher dispatch costs, due to forecast errors. At high levels of ESS, deviation of the market model from ESS technical characteristics will likely result in operational challenges, analogous to the way in which inaccurate market models of transmission and of unit commitment result in operational challenges.    

With Sections~\ref{hispersp} and~\ref{adapelect} as background, Section~\ref{impstor} then presents a stylized design for a day-ahead (DA) market, similar to that for dispatching pumped-storage hydro in the Pennsylvania-New Jersey-Maryland (PJM) system, that we believe is most compatible with our interpretation of the design philosophy of US markets. Crucially, we believe our design is compatible with high levels of ESS because it more accurately reflects ESS technological characteristics.\footnote{We acknowledge that the ratio of energy storage capacity to power capacity for typical pumped-storage hydro is typically higher than for current battery storage.} 

Section~\ref{implicload} considers demand-side resources and argues that a specialization of the proposed storage model may also facilitate the participation of demand-side resources in electricity markets.  Section~\ref{concsect} concludes. The Appendices sketch several extensions of the stylized design to illustrate their incorporation into the market model while not disrupting the main points in the paper. In particular, the more complicated case of real-time (RT) markets is sketched in the Appendices.

\section{Philosophy of US organized markets} \label{hispersp}
A useful way to view the evolution of US markets is to observe that various technical characteristics of generation and transmission have been represented explicitly into the market design, into the optimization formulation for commitment and dispatch, and into the resulting prices. These characteristics have been gradually incorporated to improve the efficiency of commitment and dispatch by reflecting economically significant technical characteristics and costs into the optimization model, with the assumption that there will generally be sufficient competition in the market for offers and bids to reveal cost information truthfully; that is, for offers to reflect marginal costs and for bids to reflect marginal willingness-to-pay.  (See, for example, \cite{KiS04,Sto02} for discussions of the relationship between competition and the revelation of costs.)

As an example of this evolution in the context of {\em transmission}, some initial US market designs, as in the Electric Reliability Council of Texas (ERCOT) did not have an explicit representation of the effect of transmission constraints. However, particularly as new generation was added in unexpected locations in ERCOT, limitations on transmission capacity became significant and actions taken out-of-market to avoid overloading capacity became economically significant and inefficient. A zonal model was then introduced in ERCOT to better represent transmission issues, but still involved costly out-of-market actions because there were significant intra-zonal transmission limitations and because the zonal model ignored the variation over locations within a zone of the effect of injections on transmission flows~\cite{Bal03a}.

Further evolution of electricity market designs then involved detailed power flow models representing all transmission lines, albeit typically approximated through ``DC power flow'' or other linearizations~\cite{WoW96}. The representation of transmission constraints implies geographically differentiated prices. Today all US organized markets use locational marginal pricing (LMP), or a variation, for payments to generators based on representation of transmission issues into dispatch and pricing. Most markets also consider linearized representations of losses. Internationally, even some markets that use zonal average prices for payments to generators, such as the Australian NEM, nevertheless dispatch based on a detailed representation of transmission constraints~\cite{AEMO24}.

As an example of this evolution in the context of {\em generation}, consider the ``offer structure,'' by which we mean the information provided to the ISO by generators that the ISO uses in its optimization formulation. For convenience, we will use offer structure to mean both:
\begin{itemize}
\item the price--quantity information, typically updated daily (with possibly different price--quantity specifications for each of the 24 hour-long intervals represented in the dispatch horizon of a DA market) or hourly (with usually a fixed price--quantity specification applying to all of the typically 12 intervals within an hour in the RT market), and
\item other technical information relating to constraints and characteristics of each generator, which is typically updated only occasionally.
\end{itemize}
Initial designs for offer structures in ERCOT and the California ISO (CAISO) consisted of a specification of simply an offer price per unit energy versus power quantity and rudimentary technical information such as minimum and maximum capacity.

The incorporation of a specification of price per unit energy versus power in the offer structure reflects that fuel costs to generate energy are the most important operating cost driver for a thermal generator, and that the efficiency of conversion of fuel into electricity can vary with power production level. Initial market designs evolved to include additional generation technical issues, including: unit commitment costs, such as start-up and no-load or min-load costs; unit commitment constraints, such as minimum up- and down-time constraints; and, ramp rate constraints. The start-up and no-load or min-load elements of the offer structure also reflect other fundamental drivers of operating costs, namely that fuel needs to be expended to bring the thermal generator to the state where it can synchronize with the electrical system and that operating at minimum generation may involve higher average operating costs than operating above minimum.

The additional details of unit commitment costs and constraints has led to transmission-constrained unit commitment market formulations by ISOs, utilizing so-called ``three-part'' offers consisting of start-up and either no-load or min-load offers, together with offer prices per unit energy for generating at power levels above minimum. That is, the offer structure consists of unit commitment costs (that is, start-up and no-load or min-load costs) and constraints together with a temporally separable offer for incremental dispatch above minimum (but without limits on energy provided over time) together with the specification of the inter-temporal constraints. Arguably, such an offer structure is a reasonable approximation to the important costs and constraints for most thermal generation. The combination of start-up, no-load, and incremental energy offers in a three-part offer is in use today in all US DA markets.  Some RT markets, such as ERCOT, do not consider unit commitment explicitly; however, other RT markets in the US also include unit commitment issues with a three-part offer.

The three-part offer representation is ``complex,'' requires the solution of mixed-integer optimization, and the non-convexity of the resulting revealed welfare maximization problem implies that LMPs for power may not cover all of the offered costs, defined to be the sum of the start-up, min-load, and the integral of the incremental energy offers evaluated at the dispatched generation levels. This has led to the establishment of side-payments, called ``make-whole'' payments, that cover any shortfall between the revenues between payment for energy and the offered cost, or more generally, provide payments to align individual incentives with the optimal commitment and dispatch. Over the last several decades, the enormous improvement in mixed integer linear programming software and of computing has enabled the incorporation of such features, while remaining within design requirements for time to solution, with a view to optimizing the efficiency of commitment and dispatch~\cite{BHH05}.

In addition to the representation of dispatched power levels, AS and other issues have also been incorporated into the commitment and dispatch model. While we will not discuss AS in detail, we observe that all of these products, including both power and AS, have been specified in terms of a minimal set of requirements that are, as far as possible, technology-neutral. For example, most AS can, in principle, be provided by any resource, whether thermal, ESS, renewable, or demand, that can satisfy the requirements. However, we emphasize that the offer structure and optimization models in place today typically include a detailed representation of the costs and operating characteristics of thermal generation and have only relatively recently begun to be expanded to represent characteristics of other generators, ESS, and demand.

To understand this evolution of the market model, it is useful to recognize that when US markets were being set up in the 1990s, much of the generation fleet in most restructured markets was thermal.  Although there were resources with close to zero marginal costs, such as nuclear power plants and some hydroelectric generation, the majority of the generation fleet typically consisted of thermal resources with non-zero marginal costs. To summarize, the market design was intended to be technology neutral but recognized the essential characteristics of the predominantly thermal generation fleet.

The increasing prevalence of renewables has necessitated changes to the market design that better reflect their technical characteristics. For example, although renewable generators have a rated capacity, the actual available capacity from a renewable generator depends on the renewable resource available to it at the time.  This renewable available capacity (RAC) therefore varies over time with the renewable resource availability. Although this seems analogous to a ``partial derate'' of a thermal generator, it is typical that partial derates for thermal generators are due to failed components of the generator itself and, moreover, they change the thermal available capacity over extended periods of time until repairs are effected.

In contrast, the RAC can vary rapidly and depends on the weather. A response to this issue in ERCOT and CAISO, for example, has been to adjust the maximum capacity for dispatch in the RT market based on telemetered values or short-term forecast of the RAC\@. That is, the offer structure for renewables for the RT market has been re-interpreted as being subject to availability of renewable resource in the short-term and this has been represented into dispatch algorithms. This feature recognizes an essential characteristic of renewable generation. That is, the introduction of significant renewable resources has resulted in changes to markets to represent the technical characteristics of renewable resources.

To summarize, products such as power and AS have been defined in US ISO markets to be technology neutral to the extent possible. However, the commitment and dispatch formulations have included specific features of generation technology such as thermal and renewable generators, and features of transmission, with a view to maximizing welfare based on models that represent the characteristics of the technologies. Offer structures have expanded to include parameters for these costs and constraints. 

For example, thermal generation incurs a fuel cost for its input fuel, with that fuel cost incurred outside of the electricity market, and then, given the fuel cost, the generator generally has a well-defined marginal cost of production when operating. Such thermal generation rarely has limitations on energy production over the hours to days timeframe. These economically significant features of thermal generation are represented in ISO/RTO models.

Turning to ESS, in addition to the pumped storage model, several ISOs began creating new participation models for batteries and flywheels in the early 2010s. Subsequently, the US Federal Energy Regulatory Commission (FERC) deliberated on this and FERC Order 841 states that:
``We find that requiring each RTO/ISO to create a participation model that recognizes the unique characteristics of electric storage resources will help eliminate barriers to their participation in the RTO/ISO markets, which will enhance competition and, in turn, help to ensure that these markets produce just and reasonable rates''~\cite[pages~41 and~42]{FERC18}. That is, FERC has ordered that markets should also adapt to ESS technical characteristics.  The next section will discuss such adaptations to markets that are needed for integration of ESS\@.

\section{Adapting electricity market design to storage} \label{adapelect}
As mentioned in the Introduction, the capacity of ESS in US markets has been growing rapidly. ESS participate in US markets today in one of two general ways.  The first, historically used by pumped-storage hydro and now typically used by chemical batteries, is by being modeled as a generator, with an offer to discharge in each interval of the dispatch horizon when it is generating, paired with a load, with a bid to charge in each interval of the dispatch horizon when it is storing energy. The ESS owner either specifies the intervals for functioning as generator and functioning as load, or allows that decision to be made pursuant to an amalgamated bid and offer.

The salient and most economically significant features of ESS are the round-trip efficiency and the limitations on energy charging and discharging due to SOC limits. In contrast to thermal generation, ESS are energy-limited and {\em do not\/} have a marginal cost of production or consumption that is independent of the electricity market.  ESS costs to generate depend on the costs to buy electricity to recharge the ESS, which inherently depend on forecasts of electricity market prices, and on round-trip efficiencies and limitations on SOC\@.

The efficient dispatch of ESS using interval-by-interval bids and offers as in most current market designs therefore requires that forecasts of prices by the ESS are accurate enough to result in correct charging and discharging decisions by the ESS\@. This current situation with ESS is analogous to that for AS before power and AS co-optimization was adopted by ISOs.  Prior to co-optimization, market participants had to forecast energy prices in order to form their AS offers, since the cost of using capacity to provide an AS is predominantly due to the opportunity cost of not using that capacity to provide power and energy. Inevitable errors in forecasts resulted in poor dispatch of capacity between power and AS.

Similarly, requiring bids to charge and offers to discharge for ESS inevitably reduces efficiency of dispatch of ESS because forming such bids and offers requires forecasts of prices.\footnote{Similarly, proposals for ``pay-as-bid'' prices in electricity markets, instead of single clearing prices, also necessitate forecasts by market participants in order to form offers, with analogous implications for reduced efficiency~\protect\cite{Bal09}.}  As discussed in~\cite{BCH23}, price forecasts by market participants are subject to at least two sources of uncertainty:
\begin{itemize}
\item due to underlying random events, including random production by weather-dependent renewables, and
\item regarding knowledge of system conditions such as availability of other generation assets, transmission constraints, and other technical issues.
\end{itemize}
As ESS become a larger part of overall operations, these errors will become increasingly significant. Moreover, the forecasting by ESS will increasingly suffer from endogeneity: the prices being forecast by an ESS need to incorporate the effect of forecasting by other ESS\@.\footnote{Other recent proposals, such as in~\cite{CAISO26,ChL25,ZQW23} that allow for SOC dependent bids and offers, and bid-ask spread models~\cite{LeJ18}\cite[Section~4]{NYISO26}, still rely on forecasting of prices to inform bids and offers and therefore are subject to similar criticisms. Appendix~\protect\ref{SOCdependent} addresses the case where charging and discharging limits depend on the SOC.}

Instead of requiring interval-by-interval bids and offers, our proposal follows and expands upon the second general approach to representation of storage as has been used in PJM for its pumped-storage hydro assets~\cite{GGG20}. In particular, the PJM approach directly models the technical characteristics of storage, including round-trip efficiency and SOC of the pumped-storage hydro, but does not include any interval-by-interval bid or offer. CAISO~\cite{CAISO23} and the New York ISO (NYISO)\cite[Section~4]{NYISO26} also have models that represent SOC, but in contrast include interval-by-interval bids and offers. None of these three ISO models include an end-of-horizon bid/offer for deviating the end-of-horizon SOC from the start-of-horizon SOC~\cite{CAISO23}\cite[Section~4]{NYISO26}.\footnote{The PJM pumped storage model does consider a constraint on final SOC~\protect\cite{GGG20} and the CAISO and NYISO models are apparently similar. We acknowledge that the PJM pumped-storage hydro model is not fully integrated into its DA model, whereas the intention in this paper is for a model of ESS, including SOC, to be fully integrated within the DA market model.}

In the context of AS, co-optimization eliminated the need for an AS offer to be based on price forecasts since opportunity costs were automatically considered in the co-optimization formulation. Similarly, representation of ESS SOC and its optimization by the ISO will eliminate the need for the ESS to forecast prices {\em within\/} the dispatch horizon to form interval-by-interval bids and offers.\footnote{An interval-by-interval term is included to represent ESS degradation in the model proposed here.} Instead, for price formation in the RT lookahead market the ISO will utilize its own forecast of renewable availabilty to resolve the first source of uncertainty mentioned above. For price formation in the DA market, that uncertainty is effectively resolved by the information in the offers submitted to the ISO by all of the market participants. In both RT and DA, the ISO will also have knowledge about system conditions, resolving the second source of uncertainty.

To summarize, the representation of storage as paired offers for generation and bids for consumption fundamentally {\em does not\/} and {\em cannot\/} represent the most salient physical and operational characteristics of storage without relying crucially on a price forecast, thereby compromising efficiency of dispatch, particularly at high levels of ESS and renewable penetration. The examples in Section~\ref{examplesect} will be used to demonstrate this observation. The salient characteristics of storage are, in fact, almost the reverse of thermal generation, since storage explicitly has inter-temporal linkages and no fuel costs. The approach presented in this paper removes entirely the need for ESS forecasts of prices within the scheduling horizon and will thereby be capable of scaling to conditions of integrating large amounts of ESS\@.\footnote{As discussed in~\protect\cite{BCH23} and Section~\protect\ref{discsection}, the ESS will still need to forecast the value of energy after the end of the horizon, either the end of the DA horizon, or the end of the RT lookahead horizon. For the DA, this value is mostly affected by longer-term supply--demand issues and weather forecasts. For the RT, such forecasts are admittedly still somewhat problematic. As an alternative to ESS bids/offers in RT, \protect\cite{BGC23} uses ISO forecasts of prices after the end of the horizon.}

In addition to round-trip efficiency and limitations on SOC, ESS may also typically exhibit additional characteristics such as ramp rate constraints and commitment decisions that are similar to the analogous characteristics of thermal generation, and these should also be represented in a market model for storage that represents SOC and round-trip efficiency. (See, for example, Appendices~\ref{mutualexcsect} and~\ref{nonzerolower} for discussion of commitment decisions for ESS.) The key point, however, is that SOC and round-trip efficiency are essential to the faithful representation of storage as demanded by FERC Order 841, despite several filings made by US ISOs that ignore this essential observation. The next section will present a model of storage that is consistent with the philosophy of representing technical characteristics into the market design and, indeed, consistent with FERC Order 841.

\section{Proposed model of ESS} \label{impstor}
The key conceptual addition for modeling ESS is the definition of variables to represent the SOC at the end of each interval together with a requirement on SOC at the end of the scheduling horizon, or more generally, a bid/offer for deviation of the end-of-horizon SOC from the beginning-of-horizon SOC\@.  This type of formulation has been utilized in longer-term pumped storage hydroelectricity models, where the ``interval'' might be a day, week, or longer, but in the DA context the interval would be an hour and in RT it would typically be 5 minutes. The end of horizon bid/offer would reflect the storage owner’s assessment of the value of buying and selling energy in the future after the end of the horizon. In the RT context, this assessment would be informed by its DA position.

A simplest model is linear with fixed round-trip efficiencies. However, the round-trip efficiency of storage devices may actually exhibit dependence on the SOC or on the charge or discharge rate, requiring a nonlinear model to represent.  A simplified approach, with fixed round-trip efficiencies, may not be able to exactly represent all the details of storage technicalities, but may be adequate for representing the salient issues, just as the models of thermal generators used in US markets are not exact, but sufficient to represent the most important technical issues.  As with the representation of thermal generation, this starting point of a highly simplified model may need to evolve as experience accumulates.

The discussion follows, with some modifications, \cite{BCH23,ChB22,BGC23,PTB11,SiE20,SDA22} and, consistent with  Section~\ref{hispersp}, intends to model the technical characteristics of a prototypical ESS\@. We assume that the dispatch horizon consists of intervals $t = 1,\ldots,T$. The interval $t=0$ will denote the interval immediately prior to the beginning of the horizon.  We will predominantly discuss DA markets so that the horizon consists of the intervals for tomorrow, but, with some modifications, the analysis also applies to RT lookahead dispatch with physically binding decisions for $t=1$ and financially binding decisions for $t = 2,\ldots,T$.\footnote{See Appendix~\protect\ref{RTmarketdef} for extension to RT markets. We also acknowledge that current RT lookahead markets do not financially bind for intervals $t = 2,\ldots,T$, resulting in various complications for the operation of these markets. See, for example, \protect\cite{ChP23,Hog21,HSZ19,May24,ZZL20} for a discussion of this and related issues. Furthermore, to implement financially binding RT decisions, a multi-settlement system would be necessary for successive RT lookahead dispatches, as described in~\cite{ZZL20}. More practical alternatives to financially binding lookahead and multi-settlements are discussed in~\protect\cite{HSZ19,ZZL20}.} We focus on one particular ESS, although in practice we expect there to be many, perhaps hundreds of ESS participating in the market that would be modeled at their individual locations in the system.

\subsection{Charging and generation variables} \label{chargdischargevarsect}
We define the charging power of the ESS at the end of interval $t$ to be $\pumpvar_{t} \in \reals$, while we define the discharging power at the end of interval $t$ to be $\genvar_{t} \in \reals$. For DA markets, the specified power level at the end of the interval will apply from just after the beginning of the interval and apply throughout the rest of the interval.  That is, generation and demand levels are assumed to ``jump'' immediately after the beginning of each interval and hold constant throughout the rest of the interval, consistent with the typical interpretation of DA markets as establishing a ``strip'' of financial forward prices for each of the RT pricing intervals within each DA interval.  Appendix~\ref{RTmarketdef} discusses the extension to RT markets where power levels are assumed to ramp within intervals.

For the simplest analysis, we assume that the ESS can charge at any value between zero and a maximum charging capacity and discharge at any value of generation between zero and a maximum generation capacity, expressed as follows:
\eab
0 \leq \pumpvar_{t} \leq \overline{\pumpvar}, & \forall t = 1,\ldots,T, &
\label{pumplimitsnobin} \\
0 \leq \genvar_{t} \leq \overline{\genvar}, & \forall t = 1,\ldots,T, &
\label{genlimitsnobin} \\
\pumpvar_{t} \perp \genvar_{t}, & \forall t = 1,\ldots,T, &
\label{pumpgenperpnobin}
\eae
where ``$\perp$'' in~(\ref{pumpgenperpnobin}) denotes that the variables on either side of it cannot both be positive. The requirement in~(\ref{pumpgenperpnobin}) is typically implemented in practice with an additional commitment binary variable; however, under some assumptions the condition will automatically be satisfied in a welfare-maximization formulation~\cite{ChB22}.\footnote{Joint dependencies on SOC and charge and discharge power will be discussed in Appendix~\ref{SOCdependent}. The extension to the case where mutual exclusivity must be explicitly enforced will be discussed in Appendix~\protect\ref{mutualexcsect}. The case of non-zero lower limits in charging and discharging will be discussed in Appendix~\protect\ref{nonzerolower}.} We collect the charging and discharging variables for $t = 1,\ldots,T$ into vectors $\pumpvar, \genvar \in \reals^{T}$.

\subsection{State-of-charge and limits} \label{soclimsect}
We define $\statevar_{t} \in \reals$ to be the SOC at the end of interval $t$.  The SOC just prior to the beginning of the horizon for tomorrow is $\statevar_{0} \in \reals$ and is assumed to be a known constant initial condition given by the end-of-horizon SOC for today.\footnote{The specification of $\statevar_{0}$ presents some problems in typical DA markets since the gate closure for submitting data to the DA market is typically the morning before the day being considered.  That is, the initial condition actually represents the SOC many hours into the future after the market formulation is solved and is therefore somewhat uncertain. However, this problem is avoided in the context of a financially binding forward market by using the financially binding SOC for the end of today.} There are conversion losses both during charging and discharging and possibly also a conversion of units between electrical energy (for example, measured in MWh) and the stored energy represented by $\statevar_{t}$ (for example, measured as mole quantity of a particular chemical species). There may also be self-discharge of the ESS\@. To model this, the SOC evolves according to:
\eb
\statevar_{t} = \change T \alpha \pumpvar_{t} - \change T \beta \genvar_{t} + \gamma \statevar_{t-1}, \forall t = 1,\ldots,T,
\label{statetransition}
\ee
where: $\change T$ is the length of the interval; the coefficients $\alpha$ and $\beta$ satisfy $\beta > \alpha > 0$ implying that the round-trip efficiency of charging and subsequently discharging is $\alpha/\beta < 1$; and, $0 < \gamma \leq 1$ represents the per interval self-discharge.\footnote{We ignore variation of round-trip efficiency and of self-discharge with SOC\@.} We collect the SOC for $t = 1,\ldots,T$ into a vector $\statevar \in \reals^{T}$; that is, $\statevar$ does not include the initial SOC, since $\statevar_{0}$ is assumed to be known and constant.

For future convenience, we define $\overline{\overline{\statevar}}_{T} = (\gamma)^T\statevar_{0}$, which is the SOC at the end of the DA horizon if the ESS were to neither charge nor discharge during the horizon; that is, if the ESS did not participate in the DA market.\footnote{In Appendix~\ref{RTSOCEOHsect}, the parameter $\overline{\overline{\statevar}}_{T}$ will be re-defined for RT markets but will correspondingly be the SOC at the end of the RT horizon if the ESS did not participate in the RT market in the sense that its RT deviations from DA were zero.} For example, if $\statevar_{0} = \mbox{100MWh}, \gamma = 0.99$, and $T = 24$ then $\overline{\overline{\statevar}}_{T} = (0.99)^{24} \times \mbox{100MWh} = \mbox{78.6MWh}$. This value is effectively the reference for considering offers and willingness-to-pay by the ESS for deviating its end-of-horizon SOC from (the implications of) its beginning-of-horizon SOC\@.

There is a maximum allowable SOC, $\overline{\statevar}$, and a minimum allowable SOC, $\underline{\statevar}$, which limit the values of $\statevar_{t}, t = 0,\ldots,T$:
\eab
\underline{\statevar} \widespace \leq \widespace \statevar_{0} & \leq & \overline{\statevar}, \label{statelimits0} \\
\gamma \statevar_{t-1}  + \change T \alpha \pumpvar_{t} &\leq& \overline{\statevar}, \forall t = 1,\ldots,T, \label{statelimits1a} \\
\gamma \statevar_{t-1}  - \change T \beta \genvar_{t} &\geq& \underline{\statevar}, \forall t = 1,\ldots,T. \label{statelimits2a}
\eae
where, for later convenience, we have separated the conditions~(\ref{statelimits0}) for the initial SOC for $t=0$ from the upper and lower bound conditions~(\ref{statelimits1a}) and~(\ref{statelimits2a}) for $t = 1,\ldots,T$. As discussed in~\cite{BCH23}, because of~(\ref{pumpgenperpnobin}), the constraints (\ref{statelimits1a})--(\ref{statelimits2a}) result in a tighter formulation than upper and lower bound constraints directly on $\statevar_{t}$.

\subsection{ESS costs and bid/offer}
Charging and discharging will typically degrade the ESS\@.  Degradation models can be complex and nonlinear; however, a simplest model is to assume that degradation during charging in interval $t$ is proportional to $\pumpvar_{t}$, while degradation during discharging in interval $t$ is proportional to $\genvar_{t}$~\cite{PBF20,Xu22}.  This results in a cost function relating to degradation for the ESS:
\eb
\epsilon \transpose{\one}\pumpvar + \zeta \transpose{\one}\genvar,
\label{ESScost}
\ee
where $\one$ is the vector of all ones, $\transpose{\mbox{ }}$ denotes transpose, and $\epsilon, \zeta \in \reals_{+}$ are coefficients representing degradation costs per unit time.\footnote{There may be a need for suitable market power mitigation regarding the coefficients $\alpha, \beta, \gamma, \epsilon$, and $\zeta$  in~(\protect\ref{statelimits1a})--(\protect\ref{ESScost}), including allowing for only occasional updates and limiting the magnitude of the values.}

As observed above, if the ESS neither charges nor discharges over the horizon then its final SOC is given by $\statevar_{T} = \overline{\overline{\statevar}}_{T}$. It is also possible for non-zero charging and discharging to occur over the horizon and for the final SOC to equal $\overline{\overline{\statevar}}_{T}$. In that case there would be degradation costs accounted for by~(\ref{ESScost}) even though $\statevar_{T} = \overline{\overline{\statevar}}_{T}$. 

Additionally, to allow for deviation of the final SOC from $\overline{\overline{\statevar}}_{T}$, we consider a bid for deviations~\cite{SMK24,SiE20}. In particular, we assume that the ESS specifies a (net) willingness-to-pay function $w: \reals \rightarrow \reals$ such that it is willing-to-pay (or receive if negative) an amount $w(\statevar_{T} - \overline{\overline{\statevar}}_{T})$ if its final SOC is equal to $\statevar_{T}$. The function $w$ is monotonically non-increasing and we define $b: \reals \rightarrow \reals$ to be the definite integral of $w$ between $0$ and $(\statevar_{T} - \overline{\overline{\statevar}}_{T})$, so that $b$ is concave and $b(0) = 0$. 

The function $b$ is the bid benefit of deviating the end-of-horizon ESS SOC from the value, $\overline{\overline{\statevar}}_{T}$, that the SOC would have been absent market participation over the horizon.\footnote{In the context of a dynamic programming formulation where we consider the SOC to be the state, $b$ is the ``value-to-go'' from the end-of-horizon.}  For positive values of its argument, the willingness-to-pay reflects the ESS owner's estimate of the value of having additional stored energy to sell after the end of the horizon, while for negative values of its argument, the willingness-to-pay reflects the ESS owner's estimate of the value of having less stored energy to sell after the end of the horizon. 

For example, if the willingness-to-pay were \$40/MWh for all charging and discharging then $b(\change\statevar_{T}) = \mbox{\$40/MWh} \times \change\statevar_{T}, \forall \change\statevar_{T}$, with interpretation that if the final SOC exceeds $\overline{\overline{\statevar}}_{T}$ by, say, $\change\statevar_{T} = \mbox{10MWh}$ then the bid benefit is \$400. In optimizing dispatch, the ISO will consider both the degradation costs in~(\ref{ESScost}) and the bid benefit $b$ of deviating the end-of-horizon ESS SOC from $\overline{\overline{\statevar}}_{T}$.

It is important to recognize that the benefit $b$ is due to accumulating a net increase in SOC over the horizon, and that this benefit is not equivalent to, for example, the value of charging in the last interval of the horizon or the value of charging in any other particular interval of the horizon. Moreover, settlement is based only on energy charged and discharged and not directly on the SOC\@. That is, there is no direct payment for SOC. 

\subsection{Balance of market}
We write $x \in \reals^{N}$ for the variables defining the rest of the market over the horizon besides the ESS under consideration. We assume that the offer net cost function for the rest of the market is of the form $\transpose{c}x$, where $c \in \reals^{N}$ is the vector of per unit time coefficients derived from offers and bids. We summarize the constraints on the rest of the market as $Cx \leq d$, including for convenience both the {\em system\/} constraints such as transmission constraints and the {\em private\/} constraints such as generation limits, and where the matrix $C$ and vector $d$ have suitable dimensions. We write power balance over the horizon in the form $\pumpvar - \genvar = Bx$, where the matrix $B$ has suitable dimensions. For notational simplicity, we ignore any binary variables required for unit commitment, but observe that binary variables can be incorporated straightforwardly into the formulation.

\subsection{Market formulation and pricing model}
Combining the features of the ESS and the balance of the market, we define the ISO problem to be:
\eb
\min_{\bsarr x \in \reals^{N}, \\ \pumpvar, \genvar, \statevar \in \reals^{T} \earr}
\left\{\left.\barr{c} \change T (\transpose{c}x) \\
\negspace \mbox{ } + \change T(\epsilon \transpose{\one}\pumpvar + \zeta \transpose{\one}\genvar) \\
\mbox{ } - b(\statevar_{T} - \overline{\overline{\statevar}}_{T}) \earr \right| \barr{c} \negspace (\ref{pumplimitsnobin})-(\ref{statelimits2a}), \\ \pumpvar - \genvar = Bx \\ Cx \leq d \earr \right\}.
\label{marketmodel}
\ee
The market model~(\ref{marketmodel}) minimizes the sum of:
\begin{itemize}
\item offered net costs of the rest of the market,
\item the degradation costs of the ESS,
\item minus the bid benefits of deviating the end-of-horizon ESS SOC from the value that the SOC would have absent market participation.
\end{itemize}
The model is convex, ignoring binary variables, and maximizes (revealed) welfare. LMPs are defined in the usual way as depending on the Lagrange multipliers of system constraints~\cite{Bal18}. There is no direct payment of ESS SOC, but instead ESS discharging is paid at the LMP and charging pays at the LMP\@. For convenience, define the ``ESS surplus'' to mean the revenue from ESS discharging plus the bid benefit of deviation of end-of-horizon SOC, minus the costs of ESS charging and of degradation. With the usual argument for convex markets, the ESS surplus is non-negative.\footnote{We acknowledge that pricing in practice can involve variations on LMPs, implying that the ESS surplus might not always be non-negative under a variant pricing rule.} If energy arbitrage is valuable then ESS surplus will be strictly positive. These characteristics will be exemplified for a small model system in Section~\ref{examplesect}.

\subsection{Discussion} \label{discsection}
It is important to note that the ESS benefits from arbitrage opportunities in the formulation developed here without having to predict the market prices within the horizon.  There is no forecast of prices within the horizon required of the ESS since the ISO {\em calculates\/} the prices that are part of the optimal dispatch over the horizon, considering the round-trip efficiency of the ESS, its degradation costs, and the bid benefits of deviating the end-of-horizon ESS SOC\@.  

The bid benefits to the ESS {\em are\/} necessarily based on a forecast by the ESS of market prices beyond the end-of-horizon. Such future prices after the end of the dispatch horizon can only be, at best, forecast imperfectly by the ISO, while the risks due to any forecast errors will generally be borne by the ESS\@. We therefore argue that such forecasts, and the risks due to forecast errors, are the legitimate and appropriate prerogative of the ESS, particularly in the DA, subject to potential market power mitigation.\footnote{We acknowledge that the argument is weaker in the RT where the ISO may have better forecasts than market participants past the end of the RT lookahead horizon.  See~\protect\cite{BGC23} for an example of such a formulation.}

The development has assumed a convex market. In the practical case that binary variables are used as part of a unit commitment formulation then the ISO problem is no longer convex, and revenue adequacy is no longer guaranteed for the same reasons as in any non-convex market. The standard approach in ISOs of make-whole payments can be used in this case, but it is to be expected that the volume of make-whole payments will be small compared the total market, as is currently the case.

\subsection{Examples} \label{examplesect}
In this section, we present a small base-case example and variants to demonstrate the properties of the proposed market model, focusing on the implications for the ESS in a system with high renewable penetration and limited thermal capacity. Demand is fixed for each interval, so that maximizing welfare is equivalent to minimizing thermal costs plus ESS degradation costs, minus the benefits of deviating the end-of-horizon ESS SOC\@. Thermal generation, renewable generation, and ESS are each represented as a stylized single aggregate.   After discussing results of each example, we will also briefly consider the implications of using interval-by-interval bids as in current market designs, instead of the proposed market model.

No reserves nor other AS are modeled and there is no unit commitment. No transmission is modeled, so there is a single locational price. A base-case and variants are considered and are solved using linear programming, with~(\ref{pumpgenperpnobin}) relaxed. In all cases considered, (\ref{pumpgenperpnobin}) was satisfied by the solution.

The following values were used for both base-case and variants:
\begin{itemize}
\item Thermal offer price \$50/MWh in all intervals, with minimum production of 0MW and maximum production of 100MW;
\item Renewable offer price \$0/MWh for all available capacity, with minimum production of 0MW and maximum production equal to available capacity; and,
\item ESS characteristics $\alpha = 0.9, \beta = 1.0, \gamma = 1, \epsilon = \mbox{\$1/MWh}, \zeta = \mbox{\$1/MWh}$, and $\underline{\statevar} = \statevar_{0} = 0 \mbox{MWh}$.
\end{itemize}
The example represents a simplified DA market with $T=6$ intervals each of duration $\change T = 4$ hours. 

The demand and the RAC in each interval is shown in Table~\ref{demrenfig}. The variation in demand is intended to represent a stylized typical daily pattern of relatively low demand from midnight to 4am (interval 1), increasing demand thereafter through noon, continued high demand from noon to 4pm (interval 4) then decreasing demand in the evening through midnight. The variation in RAC is intended to represent a stylized typical daily pattern with available capacity of onshore wind peaking sometime after midnight, low available capacity of renewables in the early morning from 4am to 8am, maximum available capacity of solar between 8am and 4pm, followed by rapid decline in available capacity of renewables until midnight. The variation of demand and RAC provides two potential opportunities per day for ESS charging and discharging. 

\begin{table}
\small
\caption{Demand and renewable available capacity (RAC).}
\label{demrenfig}
\begin{center}
\begin{tabular}{||l||r|r|r|r|r|r||}
\hline \hline
Interval $t$ &	1	&	2	&	3	&	4	&	5	&	6	\\ \hline
Demand (MW) &	100	&	200	&	300	&	300	&	200	&	100	\\
RAC (MW) &	200	&	100	&	500	&	500	&	25	&	0	\\
\hline \hline
\end{tabular}
\end{center}
\vspace{-2em}
\end{table}

\subsubsection{Base-case}
In the base-case, maximum SOC is $\overline{\statevar} = \mbox{800MWh}$ and willingness-to-pay for increasing the end-of-horizon SOC compared to $\statevar_{0} = \mbox{0MWh}$ is \$40/MWh. The base-case ESS willingness-to-pay corresponds to a forecast of future prices that are moderate for the day after the day-ahead; that is, future prices in charging intervals beyond the end-of-horizon are forecast by the ESS to be below the thermal offer price of \$50/MWh.

Base-case results are shown in Table~\ref{basecaseresfig}. Thermal generation is marginal and its offer sets the price in intervals~2 and~5. The ESS is charged and discharged in intervals~1 and~2, respectively, with state-of-charge zero at the end of interval~2. Any price between \$0/MWh (offer price of the renewable resource) and \$43.1/MWh (maximum price in interval~1 that results in storage being competitive with thermal generation in interval~2) is a market clearing price in interval~1 and any price between \$41/MWh (ESS willingness-to-pay plus the discharging degradation cost) and \$50/MWh (offer price of thermal resource) is a market clearing price in interval~6 (given the specified prices in intervals~2--5). The particular choice of \$0/MWh for market clearing price in interval~1 results in a net contribution to ESS surplus from intervals~1 and~2 of \$17,240. 

Renewable generation is not fully utilized in interval~4, driving the price to zero and reflecting the familiar phenomenon of low or zero prices in the middle of the day for markets with significant solar penetration. Price is also zero in interval~3, even though the renewables are fully deployed, because the maximum SOC limit of $\overline{\statevar} = \mbox{800MWh}$ limits the total charging in intervals~3 and~4: renewable production could be somewhat lower in interval~3 and somewhat higher in interval~4 with the same SOC results at the end of interval~4. The SOC maximum capacity couples the prices in intervals~3 and~4 in the same manner that ramping and other inter-temporal constraints couple prices when binding. Intervals~3 through~6 account for more than two-thirds of the ESS surplus of \$51,951 (with particular choice of \$41/MWh for market clearing price in interval~6), noting that the end-of-horizon SOC is unchanged from $\statevar_{0}$.\footnote{Higher choices of market clearing prices in intervals~1 and~6 would reduce the contribution to ESS surplus from intervals~1 and~2 and increase the contributions from intervals~3--6. Different market rules and different specific implementations of market clearing software will result in different choices of market clearing prices and therefore different resulting ESS surplus.} 

Now consider the case where, instead of the proposed market model, interval-by-interval bids/offers from the ESS were used by the ISO, as in current market designs where it is typical that such ESS bids/offers can vary from interval to interval. To reproduce the welfare maximizing dispatch, prices in the bids and offers from the ESS would need to be sufficiently close to the prices under the proposed market model.  

The requirements on the bid/offer prices for this case are not, however, extremely tight to achieve welfare optimality. For example, a bid/offer in each interval for an arbitrary quantity of charging or discharging at a fixed price would result in welfare maximization for any fixed price in the range from \$0/MWh to \$50/MWh. 

Such fixed price interval-by-interval bids/offers would nevertheless result in negative ESS surplus. To achieve positive surplus, interval-by-interval bids and offers would need to vary to be closer to the interval-by-interval clearing prices observed in the welfare maximizing solution. That is, the ESS would need a better forecast of the prices. As emphasized in Section~\ref{discsection}, the proposed market model provides the ESS with arbitrage profits and positive surplus without the need at all for the ESS to predict prices within the horizon.

\begin{table}
\small
\caption{Base-case results for Renewable (R), Thermal (T) and ESS (E), and Prices.}
\label{basecaseresfig}
\begin{center}
\begin{tabular}{||l||r|r|r|r|r|r||}
\hline \hline    
Interval $t$ &	1	&	2	&	3	&	4	&	5	&	6	\\ \hline
R Prod. (MW) &	200	& 100	& 500	&	322	&	25	&	0	\\
T Prod. (MW) &	0	&	10	&	0	&	0	&	75	&	0	\\
E Dis. (MW) &	0	&	90	&	0	&	0	&	100	&	100	\\
E Char. (MW) &	100	&	0	&	200	&	22	&	0	&	0	\\
E SOC (MWh) &	360	&	0	&	720	&	800	&	400	&	0	\\
Price (\$/MWh) &	0	&	50	&	0	&	0	&	50	&	41	\\
\hline \hline
\end{tabular}
\end{center}

\caption{High willingness-to-pay results.}
\label{highwtpresfig}
\begin{center}
\begin{tabular}{||l||r|r|r|r|r|r||}
\hline \hline
Interval $t$ &	1	&	2	&	3	&	4	&	5	&	6	\\ \hline
R Prod. (MW) &	200 &	100	&	500	&	322	&	25	&	0	\\
T Prod. (MW) &	0	&	10	&	0	&	0	&	100	&		100 	\\
E Dis. (MW) &	 0	&	90	&	0	&	 0 	&	75 	&	 	0 	\\
E Char. (MW) &	100	&	0 	&	200	&	22 &	0 	&		0 	\\
E SOC (MWh) &	360 	&	0	&	720 &	800 &	500 	&		 500 	\\
Price (\$/MWh) &	0 	&	50	&	 0 &	 0 	&		56 	&	 	56 	\\
\hline \hline
\end{tabular}
\end{center}

\caption{Increased maximum SOC results.}
\label{incSOCresfig}
\begin{center}
\begin{tabular}{||l||r|r|r|r|r|r||}
\hline \hline
Interval $t$ &	1	&	2	&	3	&	4	&	5	&	6	\\ \hline 
R Prod. (MW) &	200	&	100	&	500	&	 433		&	25	&	0	\\
T Prod. (MW) &	 	0 	&	 	10 	&	 	0 	&	0	&	 	0 	&		0 	\\
E Dis. (MW) &	 	0 	&	 	90 	&		0 	&	 	0 	&		175 	&	 	100 	\\
E Char. (MW) &	 	100 	&	 	0 	&		200 	&	 133		&	 	0 	&		0 	\\
E SOC (MWh) &	 	360 	&		0 	&		720 	&		\negspace 1200 	&		500 	&		100 	\\
Price (\$/MWh) &	 	0 	&		50 	&		0 	&	 	0 	&	 	41 	&	 	41 	\\
\hline \hline
\end{tabular}
\end{center}
\vspace{-2em}
\end{table}

\subsubsection{High willingness-to-pay}
In the high willingness-to-pay case, willingness-to-pay for increasing the end-of-horizon SOC compared to $\statevar_{0} = \mbox{0MWh}$ is \$55/MWh. All other parameters are the same as in the base-case. The high willingness-to-pay case corresponds to a forecast of future prices that increase for the day after the day-ahead; that is, future prices in charging intervals beyond the end-of-horizon are forecast by the ESS to be above the thermal offer price of \$50/MWh.

High willingness-to-pay case results are shown in Table~\ref{highwtpresfig}. Thermal generation is now marginal only in interval 2, and is fully dispatched in intervals~5 and~6. The ESS is charged and discharged in intervals 1 and 2, respectively, as in the base-case, with state-of-charge zero at the end of interval 2. 

Renewable production is the same as in the base-case. Thermal generation has increased significantly, allowing for a non-zero end-of-horizon ESS SOC of $\statevar_{T} = \mbox{500MWh}$. The ESS willingness-to-pay plus the discharging degradation cost set the prices in intervals 5 and 6 so that the ESS willingness-to-pay can be said to be marginal.  Intervals~3 through~6 account for nearly three-quarters of the ESS surplus of \$60,351.

Again consider interval-by-interval bids/offers. Welfare optimality would be achieved, for example, with a bid/offer in each interval for an arbitrary quantity at a fixed price in the range from \$0/MWh to \$50/MWh.  However, again, this would result in negative ESS surplus.  A better forecast of interval-by-interval prices would again be required to achieve positive ESS surplus. Interestingly, a fixed ESS bid/offer price equal to \$55/MWh in each interval would not result in welfare maximization because it would involve less renewable production and more thermal generation instead of ESS discharging in intervals~2, 5, and~6.

\subsubsection{Increased maximum SOC}
In the increased maximum SOC case, $\overline{\statevar} = \mbox{1200MWh}$. All other parameters are the same as in the base-case. Results are shown in Table~\ref{incSOCresfig}. Thermal generation is now marginal only in interval~2, and is not dispatched in any other intervals. The ESS is charged and discharged in intervals~1 and~2, respectively, as in the base-case, with state-of-charge zero at the end of interval~2.  

Increasing the maximum SOC has allowed renewable production to increase significantly in interval~4 compared to the other cases, both displacing thermal generation and allowing for non-zero end-of-horizon ESS SOC of $\statevar_{T} = \mbox{100MWh}$. The ESS willingness-to-pay plus the discharging degradation cost again set the prices in intervals~5 and~6 and this has lowered the price compared to the base-case in interval~5.  Intervals~3 through~6 again account for most of the ESS surplus of \$63,907.

Again consider interval-by-interval bids/offers. Achieving positive surplus would again require a forecast of interval-by-interval prices. 

\subsection{Large-scale implementation}
The development and examples have involved a single ESS for expositional clarity. However, there may be hundreds of large ESS in future systems. This section discusses the computational implications of large-scale implementation.

Computationally, representing SOC involves constraints that couple decisions across all intervals in the horizon.  For example, in a DA market with 24 hourly intervals, there will be SOC coupling constraints between each hour that couple all hours together.  This increases the density of the constraint matrices in the optimization problem solved by the ISO and inherently couples the decisions from interval to interval, potentially making the associated optimization problem too difficult to solve within required time limitations.  Recent work has demonstrated that representation of well over 100 ESS using a formulation similar to this paper is feasible for even a large system such as the Midcontinent ISO within time requirements for DA solution~\cite{BCH23,ChB22}. That is, the computational issues are surmountable in realistically large systems.

\section{Price responsive demand} \label{implicload}
In the examples in Section~\ref{examplesect}, demand is fixed. However, some markets allow for price responsive demand, typically represented with an assumed temporally separable willingness-to-pay.  For some loads, such as crypto\-currency mining, this is appropriate since the value of expending effort on mining in a particular pricing interval is either lower or higher than the price of the ``mined'' cryptocurrency and the miner can interrupt or slow computations without incurring the analog of start-up or shut-down costs.

However, in contrast, a significant fraction of other types of consumption, both industrial and residential, is akin to storage in that the consumption is not tied to particular times, but can be shifted across time, within limits, so long as sufficient service is provided over a required time horizon.  That is, over a time horizon such as a day, there are likely requirements on total energy to satisfy production commitments.  If these daily limits can be satisfied then production can potentially be flexibly scheduled within, for example, ramping limits and other constraints.

A key observation is that a specialization of the SOC model for storage can accommodate a requirement on daily energy for loads.  The implication is that a key to unlocking demand flexibility is the representation of SOC for not only conventional storage devices but also for consumers that have a daily energy requirement. In particular, by specifying the maximum discharging capacity to be zero, the degradation cost coefficients to be zero, and the self-discharge coefficient $\gamma = 1$, the model we have described for ESS can be utilized to represent a load with a willingness-to-pay for total energy consumed over a horizon. An example of such a formulation for a chemical plant is provided in~\cite{TPB26}.\footnote{Analogously, energy-limited thermal generation can also be represented by specializing the SOC model with maximum charging capacity zero, $\beta$ equal to the inverse of the heat rate, degradation cost coefficients zero, and self-discharge coefficient $\gamma = 1$. The end-of-horizon offer then represents the fuel cost and availability.}

\section{Conclusion} \label{concsect}
This paper has described a model for ESS participation in electricity markets that explicitly represents and optimizes ESS SOC considering round-trip efficiency and other technical characteristics of ESS as required by FERC Order 841~\cite[page~1]{FERC18}.  This contrasts with most filings by US ISOs in response to Order 841, which do not represent such salient technical characteristics of ESS and require forecasts by the ESS of market prices. Implementation of the model described in this paper would, however, require considerable changes to the structure of existing electricity markets. Furthermore, several issues remain including the design of AS markets. These issues will be considered in a future paper.

\section*{Acknowledgment}
The author would like to thank Yonghong Chen of the National Laboratory of the Rockies, Sergio Due\~{n}as Mel\'{e}ndez, Anna McKenna, and Mark Rothleder of CAISO, Pengwei Du and David Maggio of ERCOT, Anthony Giacomoni of PJM, Paul Gribik of Paul Gribik Consulting, Nongchao Guo of NYISO, Udi Helman of Helman Analytics, William B. Hogan of Harvard University, Bing Huang and Congcong Wang of MISO, Nicholas D. Laws of Camus Energy, Richard P. O'Neill formerly at FERC, Nikita Singhal of EPRI, Ramteen Sioshansi of CMU, Alva Svoboda of Pacific Gas \& Electric, Bolun Xu of Columbia University, and Tongxin Zheng of ISO-NE for comments on earlier drafts of the paper.

\bibliography{../../papers/bibliography}

\begin{thebibliography}{10}

\bibitem{Bal03a}
Ross Baldick.
\newblock Shift factors in {ERCOT} congestion pricing.
\newblock Technical report. Available from users.ece.utexas.edu/\~{
  }baldick/papers/shiftfactors.pdf. Accessed December 21, 2025, March 2003.

\bibitem{Bal09}
Ross Baldick.
\newblock Single clearing price in electricity markets.
\newblock Available from
  users.ece.utexas.edu/\~{}baldick/papers/baldick-single-price-auction.pdf.
  Accessed January 30, 2026, February 2009.

\bibitem{Bal18}
Ross Baldick.
\newblock Offer-based economic dispatch.
\newblock Course notes for University of Texas at Austin Course EE394V
  ``Restructured Electricity Markets: Locational Marginal Pricing.'' Available
  from users.ece.utexas.edu/\~{ }baldick/classes/394V/Offer.pdf. Accessed
  December 21, 2025, 2018.

\bibitem{Bal21}
Ross Baldick.
\newblock {CAISO} storage forum.
\newblock Presented at the CAISO Storage Forum: Energy Markets for the Future,
  October 28, 2021.

\bibitem{BCH23}
Ross Baldick, Yonghong Chen, and Bing Huang.
\newblock Optimization formulations for storage devices with disjoint operating
  modes.
\newblock {\em Operations Research}, 71(6):1978--1996, November--December 2023.

\bibitem{BHH05}
Ross Baldick, Udi Helman, Benjamin~F. Hobbs, and Richard~P. O'Neill.
\newblock Design of efficient generation markets.
\newblock {\em Proceedings of the IEEE}, 93(11):1998--2012, November 2005.

\bibitem{CAISO20}
{California ISO}.
\newblock Opinion on energy storage and distributed energy resources phase 4.
\newblock Available from
  https://www.caiso.com/documents/msc-opiniononenergystorageanddistributedresourcesphase4-sep8\_2020.pdf.
  Accessed April 15, 2026, 2020.

\bibitem{CAISO23}
{California ISO}.
\newblock Energy storage enhancements: State of charge implementation--update.
\newblock Available from
  stakeholdercenter.caiso.com/InitiativeDocuments/Workshop-Paper-Energy-Storage-Enhancements-State-of-Charge-Implementation-Update-Oct-2-2023.pdf.
  Accessed February 5, 2026, 2023.

\bibitem{CAISO26}
{California ISO}.
\newblock Storage design and modeling.
\newblock Presentation. Available from
  stakeholdercenter.caiso.com/InitiativeDocuments/Presentation-Storage-Design-Modeling-Jan-22-2026.pdf.
  Accessed January 28, 2026.

\bibitem{ChL25}
Cong Chen and Siying Li.
\newblock Multi-interval energy-reserve co-optimization with {SoC}-dependent
  bids from battery storage.
\newblock {\em IEEE Transactions on Power Systems}, 40(4):3008--3016, July
  2025.

\bibitem{ChB22}
Yonghong Chen and Ross Baldick.
\newblock Battery storage formulation and impact on day ahead security
  constrained unit commitment.
\newblock {\em IEEE Transactions on Power Systems}, 37(5):3995--4005, September
  2022.

\bibitem{ChP23}
Jehum Cho and Anthony Papavasiliou.
\newblock Pricing under uncertainty in multi-interval real-time markets.
\newblock {\em Operations Research}, 71(6):1928--1942, November--December 2023.

\bibitem{FERC18}
Federal Energy~Regulatory Commission.
\newblock Electric storage participation in markets operated by regional
  transmission organizations and independent system operators, February 2018.
\newblock Docket Nos. RM16-23-000; AD16-20-000; Order No. 841. Available from
  www.ferc.gov. Accessed October 2, 2025.

\bibitem{ENK14}
Madeleine Ecker, Nerea Nieto, Stefan {K\"{a}bitz}, et~al.
\newblock Calendar and cycle life study of {Li(NiMnCo)$\mbox{O}_{2}$}-based
  18650 lithium-ion batteries.
\newblock {\em Journal of Power Sources}, 248:839--851, February 2014.

\bibitem{GGG20}
Anthony Giacomoni, Qun Gu, and Boris Gisin.
\newblock Optimizing hydroelectric pumped storage in {PJM}'s day-ahead energy
  market.
\newblock FERC Technical Conference. Available from
  https://www.ferc.gov/sites/default/{f}{i}les/2020-06/T2-3\_Giacomoni\_et\_al.pdf.
  Accessed March 8, 2022, June 2020.

\bibitem{Hog21}
William~W. Hogan.
\newblock Electricity market design: Multi-interval pricing models.
\newblock Technical report, John F. Kennedy School of Government, Harvard
  University, July 2021.
\newblock Available from whogan.scholars.harvard.edu/file\_url/359. Accessed
  December 30, 2025.

\bibitem{HSZ19}
Bowen Hua, Dane Schiro, Tongxin Zheng, Ross Baldick, and Eugene Litvinov.
\newblock Pricing in multi-interval real-time markets.
\newblock {\em IEEE Transactions on Power Systems}, 34(4):2696--2705, July
  2019.

\bibitem{BGC23}
Bing Huang, Arezou Ghesmati, Yonghong Chen, and Ross Baldick.
\newblock A computationally efficient pumped storage hydro optimization in the
  look-ahead unit commitment and real-time market dispatch under uncertainty.
\newblock {\em Electric Power Systems Research}, 225:109798, December 2023.

\bibitem{JPM25}
{J P Morgan Private Bank}.
\newblock Heliocentrism: Objects may be further away than they appear, March
  2025.
\newblock Available from
  privatebank.jpmorgan.com/apac/en/insights/latest-and-featured/eotm/annual-energy-paper.
  Accessed November 21, 2025.

\bibitem{KiS04}
Daniel~S. Kirschen and Goran Strbac.
\newblock {\em Fundamentals of Power System Economics}.
\newblock John Wiley and Sons, Chichester, England, 2004.

\bibitem{AEMO24}
Daniel Lavis.
\newblock Guide to mis-pricing information.
\newblock Technical report, AEMO Electricity Market Monitoring, June 2024.
\newblock Available from
  www.aemo.com.au/-/media/files/electricity/nem/security\_and\_reliability/dispatch/
  policy\_and\_process/guide-to-mis-pricing-information.pdf. Accessed December
  29, 2025.

\bibitem{LeJ18}
Whitney Lesnicki and Pallavi Jain.
\newblock {ESR} participation model: Energy market design.
\newblock Available from www.nyiso.com. Accessed December 8, 2025, September
  2018.

\bibitem{May24}
Jacob Mays.
\newblock Sequential pricing of electricity.
\newblock {\em Energy Economics}, 137:107790, 2024.

\bibitem{NYISO26}
{New York ISO}.
\newblock Market services tariff.
\newblock Available from www.nyiso.com/regulatory-viewer. Accessed January 6,
  2025.

\bibitem{PTB11}
Panagiotis Patrinos, Sergio Trimboli, and Alberto Bemporad.
\newblock Stochastic {MPC} for real-time market-based optimal power dispatch.
\newblock In {\em Proceedings of the 2011 50th {IEEE} Conference on Decision
  and Control and European Control Conference (CDC-ECC)}, Orlando, FL, December
  2011. IEEE.

\bibitem{PBF20}
Yuliya Preger, Heather~M. Barkholtz, Armando Fresquez, Daniel~L. Campbell,
  et~al.
\newblock Degradation of commercial lithium-ion cells as a function of
  chemistry and cycling conditions.
\newblock {\em Journal of The Electrochemical Society}, 167(12):120532,
  September 2023.

\bibitem{QDS23}
Ming Qu, Tao Ding, Yuge Sun, et~al.
\newblock Convex hull model for a single-unit commitment problem with pumped
  hydro storage unit.
\newblock {\em IEEE Transactions on Power Systems}, 38(5):4867--4880, September
  2023.

\bibitem{SMK24}
Nikita Singhal, Alexandre Moreira, Julie~Mulvaney Kemp, et~al.
\newblock Navigating modeling frontiers for electric storage resources in
  wholesale electricity markets.
\newblock Technical report, Lawrence Berkeley National Laboratory, October
  2024.
\newblock Available from
  eta-publications.lbl.gov/sites/default/files/2024-10/navigating\_modeling\_frontiers\_for\_esrs\_vfinal\_doerev\_final.pdf.
  Accessed December 22, 2025.

\bibitem{SiE20}
Nikita~G. Singhal and Erik~G. Ela.
\newblock Pricing impacts of state of charge management options for electric
  storage resources.
\newblock In {\em 2020 IEEE Power \& Energy Society General Meeting (PESGM)},
  pages 1--6, 2020.

\bibitem{SDA22}
Ramteen Sioshansi, Paul Denholm, Juan Arteaga, et~al.
\newblock Energy-storage modeling: State-of-the-art and future research
  directions.
\newblock {\em IEEE Transactions on Power Systems}, 37(2):860--875, March 2022.

\bibitem{Sto02}
Steven Stoft.
\newblock {\em Power System Economics: Designing Markets for Electricity}.
\newblock IEEE Press and Wiley Interscience and John Wiley \& Sons, Inc.,
  Piscataway, NJ, 2002.

\bibitem{TPB26}
Xin Tang, Cosmin~G. Petra, Michael Baldea, and Ross Baldick.
\newblock A grid-scale study of demand bidding by large industrial users.
\newblock {\em Computers and Chemical Engineering}, 205:109442, 2026.

\bibitem{EIA25}
{US Energy Information Administration}.
\newblock Utility-scale batteries are more commonly used for price arbitrage,
  September 2025.
\newblock Available from www.eia.gov/todayinenergy/detail.php?id=66164.
  Accessed November 21, 2025.

\bibitem{WoW96}
Allen~J. Wood and Bruce~F. Wollenberg.
\newblock {\em Power Generation, Operation, and Control}.
\newblock Wiley, New York, second edition, 1996.

\bibitem{Xu22}
Bolun Xu.
\newblock The role of modeling battery degradation in bulk power system
  optimizations.
\newblock {\em MRS Energy \& Sustainability 9}, 9:198--211, September 2022.

\bibitem{YAS25}
Bo~Yuan, Waleed Aslam, Nikita~G. Singhal, and Robert Entriken.
\newblock Rethinking ancillary services markets: Balancing speed and endurance
  in the era of limited energy resources.
\newblock {\em IEEE Transactions on Energy Markets, Policy and Regulation},
  pages 1--15, 2025.

\bibitem{ZZL20}
Jinye Zhao, Tongxin Zheng, and Eugene Litvinov.
\newblock A multi-period market design for markets with intertemporal
  constraints.
\newblock {\em IEEE Transactions on Power Systems}, 35(4):3015--3025, July
  2020.

\bibitem{ZQW23}
N.~Zheng, X.~Qin, D.~Wu, G.~Murtaugh, and B.~Xu.
\newblock Energy storage state-of-charge market model.
\newblock {\em IEEE Transactions on Energy Markets, Policy and Regulation},
  1(1):11--22, March 2023.

\end{thebibliography}

\appendix
\section{Appendix} \label{appdetails}
Several extensions to the basic model from Section~\ref{impstor} are discussed in the following sections.
\subsection{Joint dependencies on SOC and charge and discharge power} \label{SOCdependent}
Explicit representation of SOC in the market model allows for some characteristics of ESS to be straightforwardly represented that would otherwise be problematic to represent if only power levels (or only SOC levels) are modeled. The following sections discuss two examples involving constraints and the objective, respectively.
\subsubsection{Constraints} 
At high and low levels of SOC, the allowable discharging and charging power levels may be restricted~\cite[Section 3.1.1]{YAS25}.\footnote{This is sometimes referred to as ``foldback''~\protect\cite{CAISO26}.}  Figure~\ref{jointSOCpowerfig} shows an example of such constraints, where the feasible SOC and power levels are on the interior and edge of the irregular hexagon. At low levels of $\statevar_{t}$ in interval $t$ that are close to $\underline{\statevar}$, the discharge power $\genvar_{t}$ is limited to being below $\overline{\genvar}$, while for high levels of $\statevar_{t}$ in interval $t$ that are close to $\overline{\statevar}$, the charge power $\pumpvar_{t}$ is limited to being below $\overline{\pumpvar}$. The explicit representation of SOC and power generation in the formulation allows for such requirements to be included straightforwardly using~(\ref{pumplimitsnobin})--(\ref{genlimitsnobin}) together with the two additional linear constraints shown as thick lines in Figure~\ref{jointSOCpowerfig}. In contrast, in the absence of an explicit representation of SOC and discharge and charge power, it is not possible to represent this type of restriction exactly. 

\begin{figure}
\setlength{\unitlength}{0.3\linewidth}
\begin{picture}(1.3,1.3)(-0.15,0.0)
\thinlines
\put(0.0,-0.1){\vector(1,0){1.1}}
\put(0.0,-0.1){\vector(0,1){1.1}}
\put(-0.05,1.1){$\statevar_{t}$}
\put(1.15,-0.14){$\genvar_{t} - \pumpvar_{t}$}
\put(0.2,-0.15){\line(0,1){0.1}}
\put(0.5,-0.15){\line(0,1){0.1}}
\put(0.8,-0.15){\line(0,1){0.1}}
\put(0.1,-0.27){$-\overline{\pumpvar}$}
\put(0.46,-0.27){$0$}
\put(0.76,-0.27){$\overline{\genvar}$}
\put(-0.05,0.2){\line(1,0){0.1}}
\put(-0.05,0.8){\line(1,0){0.1}}
\put(-0.15,0.17){$\underline{\statevar}$}
\put(-0.15,0.75){$\overline{\statevar}$}

\thicklines
\put(0.2,0.2){\line(1,0){0.4}}
\put(0.2,0.2){\line(0,1){0.5}}
\put(0.8,0.8){\line(-1,0){0.4}}
\put(0.8,0.8){\line(0,-1){0.5}}
\put(0.6,0.2){\line(2,1){0.2}}
\put(0.2,0.7){\line(2,1){0.2}}
\put(0.6,0.205){\line(2,1){0.2}}
\put(0.2,0.705){\line(2,1){0.2}}
\put(0.6,0.21){\line(2,1){0.2}}
\put(0.2,0.71){\line(2,1){0.2}}
\put(0.6,0.19){\line(2,1){0.2}}
\put(0.2,0.69){\line(2,1){0.2}}
\put(0.6,0.195){\line(2,1){0.2}}
\put(0.2,0.695){\line(2,1){0.2}}

\end{picture}
\hfill
\parbox{0.4\linewidth}{
\caption
{Joint constraints on SOC and power level. The feasible SOC and power levels are on the interior and edge of the irregular hexagon.}
\label{jointSOCpowerfig}
}
\end{figure}
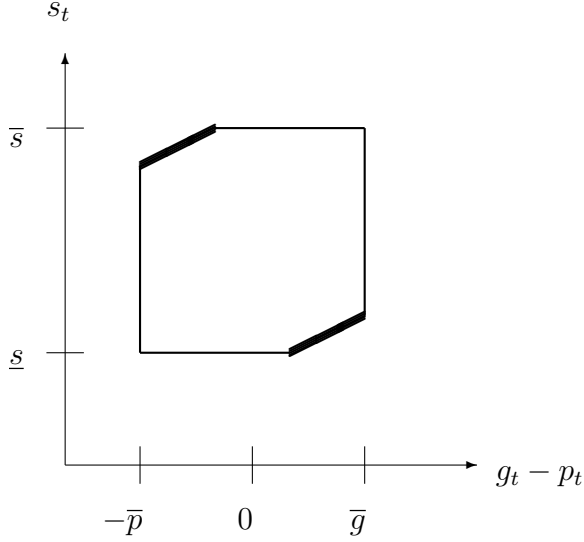

\subsubsection{Objective}
Degradation costs may depend on both SOC and power levels.  For example, \cite[Figure~12(a)]{ENK14} shows much faster degradation in capacity when a battery is cycled around values of SOC close to $\underline{\statevar}$ or $\overline{\statevar}$ as compared to cycling for values of SOC close to midpoint values of SOC $(\underline{\statevar} + \overline{\statevar})/2$. This qualitative characteristic can be approximated with a linear model using additional constraints. For example, the term for degradation due to charging in interval $t$ in~(\ref{ESScost}), $\epsilon\pumpvar_{t}$, can be replaced by the following:
\eabn
D_{t} &\geq& \epsilon\pumpvar_{t}, \\
D_{t} &\geq& \epsilon^{+}\pumpvar_{t} + (\epsilon^{+} - \epsilon)2\overline{\pumpvar}(\statevar - \overline{\statevar})/(\overline{\statevar}-\underline{\statevar}),
\eaen 
where $D_{t}$ is the contribution to degradation costs due to pumping in interval $t$; $\epsilon$ is the degradation cost for charging with SOC around $(\underline{\statevar} + \overline{\statevar})/2$; and, $\epsilon^{+}$ is the higher degradation cost for charging with SOC around $\overline{\statevar}$.  For $\statevar = (\underline{\statevar} + \overline{\statevar})/2$, the contribution is $\epsilon\pumpvar_{t}$, whereas for $\statevar = \overline{\statevar}$, the contribution is $\epsilon^{+}\pumpvar_{t}$, with intermediate contributions at intermediate values of SOC.

\subsection{Enforcing mutual exclusivity of charging and discharging} \label{mutualexcsect}
As discussed in Section~\ref{chargdischargevarsect}, under some conditions, the mutual exclusivity condition~(\ref{pumpgenperpnobin}) will be automatically satisfied in a welfare-maximization formulation~\cite{ChB22}. However, in the case that the conditions are not satisfied, one additional commitment binary variable $\pumpcvar_{t} \in \{0,1\}$ can be added per ESS per interval $t$ to enforce mutual exclusivity. In particular, following~\cite{ChB22}, (\ref{pumplimitsnobin})--(\ref{pumpgenperpnobin}) become:
\eab
0 \leq \pumpvar_{t} \leq& \negspace\negspace\pumpcvar_{t}\overline{\pumpvar}, \widespace & \forall t = 1,\ldots,T, \label{onebinary1} \\
0 \leq \genvar_{t} \leq& \negspace\negspace\negspace(1-\pumpcvar_{t})\overline{\genvar},\negspace\negspace\negspace & \forall t = 1,\ldots,T, \label{onebinary2} \\
\pumpcvar_{t} \in & \negspace\{0,1\}, \widespace & \forall t = 1,\ldots,T. \non
\eae
The binary variable $\pumpcvar_{t}$ represents whether or not the ESS is charging in interval $t$.

\subsection{Non-zero limits on charging and discharging} \label{nonzerolower}
Some storage devices have disjoint operating modes due to non-zero lower charging or discharging limits.  Representation of this typically requires two commitment binary variables for each interval.\footnote{If either the lower charging limit is zero or the lower discharging limit is zero then only one binary is necessary as in Appendix~\protect\ref{mutualexcsect}, by changing the lower bound constraint in either~(\protect\ref{onebinary1}) or~(\protect\ref{onebinary2}) to being non-zero~\protect\cite{QDS23}.}  Following~\cite{BCH23}, for each $t = 1,\ldots,T$, in addition to the binary commitment variable $\pumpcvar_{t} \in \{0,1\}$ defined in Appendix~\ref{mutualexcsect}, we define another binary commitment variable $\gencvar_{t} \in \{0,1\}$ to represent whether the ESS is discharging or not. This enables~(\ref{pumplimitsnobin})--(\ref{pumpgenperpnobin}) to be generalized to:
\eabn
\underline{\pumpvar}\pumpcvar_{t} \leq \pumpvar_{t} \leq \overline{\pumpvar}\pumpcvar_{t}, & \forall t = 1,\ldots,T, & \\
\underline{\genvar}\gencvar_{t} \leq \genvar_{t} \leq \overline{\genvar}\gencvar_{t}, & \forall t = 1,\ldots,T, & \\
\pumpcvar_{t} + \gencvar_{t} \leq 1, & \forall t = 1,\ldots,T, \\
\pumpcvar_{t} \in \{0,1\}, & \forall t = 1,\ldots,T, & \\
\gencvar_{t} \in \{0,1\}, & \forall t = 1,\ldots,T, &
\eaen
where $\underline{\pumpvar}$ and $\underline{\genvar}$ are the non-zero lower charging and discharging limits, respectively.

\subsection{Real-time markets} \label{RTmarketdef}
In RT, the ESS will have a financial position from the DA\@. RT settlements are based on deviations from the DA market, with details discussed in the following sections. 
\subsubsection{SOC evolution}
The forward net power position $(\genvar_{t} - \pumpvar_{t})$ from the DA market is assumed constant throughout each individual DA interval $t$ as embodied in the SOC update~(\ref{statetransition}). In contrast, in RT markets, power levels are assumed to ramp within intervals. For example, in ERCOT, for dispatched discharge levels of $\genvar_{t-1}$ and $\genvar_{t}$ in two successive RT intervals $t-1$ and $t$, respectively, the ESS is expected to be discharging at $\genvar_{t-1}$ at the beginning of interval $t$, linearly ramp production to $\genvar_{t}$ over the first four minutes of interval $t$, and then remain discharging at $\genvar_{t}$ for the last minute of interval $t$.\footnote{The constant dispatch level for the last minute of the interval is intended to facilitate state estimation so that new dispatch can be evaluated for interval $t$ before the end of interval $t-1$.} As discussed in~\cite[Section 4.1.3]{SMK24} and~\cite{YAS25}, this requires adjustment to the SOC update~(\ref{statetransition}) for interval $t$ because the dispatched charging and discharging from interval $t-1$ affect the SOC at the end of interval $t$. 

Transition from charging to discharging within an interval or vice versa complicates the SOC update because the net effect on SOC requires careful consideration of the transition. We avoid this complication by additional constraints on the binaries that prevent a transition {\em within\/} any interval. This results in the following SOC update and constraints on $\pumpcvar_{t-1}, \gencvar_{t-1}, \gencvar_{t}$, and $\pumpcvar_{t}$ in addition to the constraints defined in Appendix~\ref{nonzerolower}:
\eabn
\statevar_{t} &=& \change T \alpha (\eta\pumpvar_{t} + (1-\eta)\pumpvar_{t-1}) \\
& & \mbox{ }  - \change T \beta (\eta\genvar_{t} + (1-\eta)\genvar_{t-1}) + \gamma \statevar_{t-1}, \\
& &  \widespace\widespace \forall t = 1,\ldots,T, \\
\pumpcvar_{t-1} + \gencvar_{t} &\leq& 1, \forall t = 1,\ldots,T, \\
\gencvar_{t-1} + \pumpcvar_{t} &\leq& 1, \forall t = 1,\ldots,T,
\eaen
where $\eta \in [0,1]$ is the fractional contribution of the dispatch level at the end of interval $t$ to the change in SOC. For example, for ERCOT, $\eta = 3/5$, while if the ramp was linear throughout the interval then $\eta = 0.5$.

\subsubsection{ESS bid/offer} \label{RTSOCEOHsect}
To conveniently treat the RT implications of the DA financial position, consider a particular RT lookahead dispatch run and the corresponding RT end-of-horizon. If the RT end-of-horizon happens to coincide with the end of a DA interval, say DA interval $\tau$, then we re-define $\overline{\overline{\statevar}}_{T}$ for the RT market to be the DA SOC at the end of interval $\tau$.  If the RT end-of-horizon lies within DA interval $\tau$ then, consistent with the interpretation of the DA forward position as a strip, we re-define $\overline{\overline{\statevar}}_{T}$ for the RT market to be the interpolation of the corresponding DA SOCs at the beginning and end of DA interval $\tau$. We then interpret an ESS bid/offer for the RT market to be for deviations of the RT end-of-horizon SOC with respect to the re-defined $\overline{\overline{\statevar}}_{T}$. This allows for a single RT ESS deviation bid/offer for the RT to be used for multiple successive RT lookahead dispatches. For example, ESS deviation bid/offers could be adjusted hourly, with a given ESS deviation bid/offer used in 12 successive lookahead dispatches.

\subsection{Prudential issues}
In either the DA or RT markets, an ESS could have a final SOC $\statevar_{T}$ that differs from $\overline{\overline{\statevar}}_{T}$ (as defined for DA in Section~\ref{soclimsect} to be 
$\overline{\overline{\statevar}}_{T} = (\gamma)^T\statevar_{0}$ and as defined for RT in Appendix~\ref{RTSOCEOHsect} to be the interpolation of the DA SOCs). If $\statevar_{T} > \overline{\overline{\statevar}}_{T}$ in the DA then the ESS has net purchased DA energy over the DA horizon and might need to post surety based on the willingness-to-pay to ensure that it could settle these DA net purchases. Similarly, if $\statevar_{T} > \overline{\overline{\statevar}}_{T}$ in the RT then the ESS has net purchased RT energy over an RT horizon and might again need to post surety based on the willingness-to-pay to ensure that it could settle these RT net purchases. Both of these concerns would only apply for extremely high values of willingness-to-pay. 

A potentially more significant risk applies in the RT if $\statevar_{T} < \overline{\overline{\statevar}}_{T}$ because of the implications for later DA intervals. For example, if $\statevar_{T}$ is significantly below $\overline{\overline{\statevar}}_{T}$ in the RT then it might be the case that the forward positions in later DA intervals could not be fulfilled without further purchases in the RT\@.  This could happen if the rest of the DA positions involved significant discharging. 

To summarize, large deviations of RT SOC below the corresponding DA SOC entail risks of enforced adverse purchases in the RT by the ESS during the balance of the day. It might be appropriate in this case to also require posting of financial surety by the ESS for choices of bid willingness-to-pay that required purchases from the RT market to fulfill its later DA positions. 

\end{document}